# QNMnonreciprocal_resonators: an openly available toolbox for computing the QuasiNormal Modes of nonreciprocal resonators

Tong Wu, Philippe Lalanne
LP2N, Institut d’Optique d’Aquitaine, IOGS, Univ. Bordeaux, CNRS

wutong1121@sina.com
philippe.lalanne@institutoptique.fr

Last revision: June, 2021

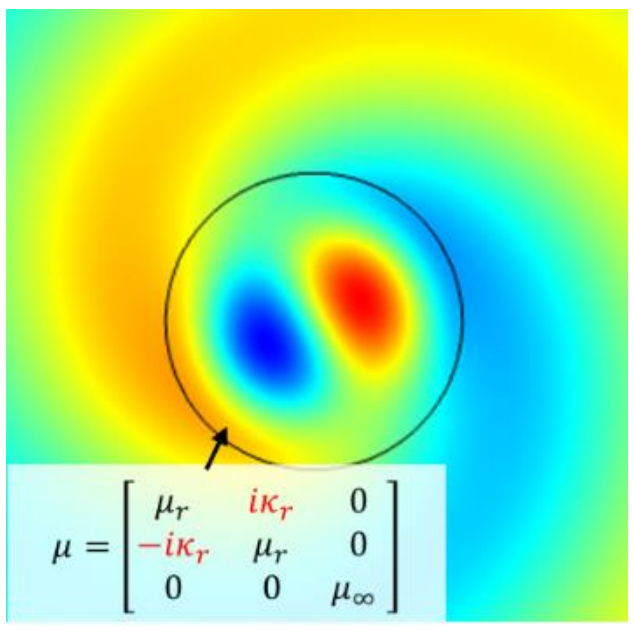

- the present user guide document,
- QNM_Dolmen_multipole.mph(.m), a COMSOL model provided under. mph or .m extensions.

# Table of contents

# 1. QNMnonreciprocal_resonators

**QNMnonreciprocal_resonators** is an extension (posted in 2021) of the **QNMEig** solver of the freeware package **MAN.** It provides a comprehensive presentation of the computation and normalization of electromagnetic quasinormal modes (QNMs) of resonators composed of nonreciprocal materials. It features a theoretical background on the topic and a COMSOL model that illustrates how to put into practice the theory on the example of a Yttrium iron garnet wire in a homogenous background. This document provides the necessary details on how the model is built so that the interested readers may easily modify it for computing QNMs of other nonreciprocal resonators.

The present model has been successfully tested with another COMSOL model operating with **QNMpole**, the second QNM solver of **MAN** that performs a pole search gradient descent algorithm to compute and normalize QNMs [1]. The corresponding COMSOL model and Matlab programs used in the **QNMpole** are also included in the version 7.2 of MAN and following ones.

## 1.1 How to acknowledge and cite

We kindly ask that you reference the **MAN** package from IOGS-CNRS and its authors in any publication/report for which you used it. The preferred citation for **QNMnonreciprocal_resonators** is the following paper:

Wu, T.; Gurioli, M.; Lalanne, P. "Nanoscale Light Confinement: the Q's and V's", *ACS Photonics* **2021**, (https://doi.org/10.1021/acsphotonics.1c00336).

## 1.2 Units and conventions of input/output data for QNMtoolbox_multipole

**Unit**. All the input information is required to be in the **SI unit**. Accordingly, the output information is given in the SI unit as well.

**Convention**. The time-dependent terms $\exp(i\omega t)$ are used in the COMSOL model provided in this toolbox. We use $\exp(-i\omega t)$ convention for the formulas in Sections 1.3 and 2.

## 1.3 Outline of the theoretical background

The Section is organized as follows. It starts with the classical derivation of a general form of the Lorentz reciprocity theorem. The theorem is then used to normalize the modes and introduce the definition of the mode volume. Unlike our previous work [2,3], we do not assume reciprocity, namely, the formulas are valid for systems with $\bar{\bar{\mu}}^T \neq \bar{\bar{\mu}}$ or $\bar{\bar{\varepsilon}}^T \neq \bar{\bar{\varepsilon}}$ (the uperspcript $T$ being the transpose operator). The theory presented here can be viewed as an extension of Supplementary Material 1 of ref. [2]. Related works may also be found in refs. [4,5].

### 1.3.1 Unconjugated form of Lorentz reciprocity theorem for nonreciprocal systems

We consider systems satisfying the following Maxwell's equations:

$$\nabla \times \mathbf{E}_i = i\omega_i \bar{\bar{\mu}}_i \mathbf{H}_i$$
$$\nabla \times \mathbf{H}_i = -i\omega_i \bar{\bar{\varepsilon}}_i \mathbf{E}_i - i\omega_i \mathbf{p}_i, \quad (1)$$

which features a source term $-i\omega_i \mathbf{p}_i$.

Lorentz reciprocity theorem relates two different solutions of Maxwell's equations, ($\mathbf{E}_1$, $\mathbf{H}_1$, $\omega_1$, $\mathbf{p}_1$, $\bar{\bar{\varepsilon}}_1$, $\bar{\bar{\mu}}_1$) and ($\mathbf{E}_2$, $\mathbf{H}_2$, $\omega_2$, $\mathbf{p}_2$, $\bar{\bar{\varepsilon}}_2$, $\bar{\bar{\mu}}_2$) labeled by indices 1 and 2. It is derived by applying the divergence theorem to the vector $\mathbf{E}_2 \times \mathbf{H}_1 - \mathbf{E}_1 \times \mathbf{H}_2$ and by using Eq. (1),

$$\iint_\Sigma (\mathbf{E}_2 \times \mathbf{H}_1 - \mathbf{E}_1 \times \mathbf{H}_2) \cdot d\mathbf{S} = i \iiint_\Omega \mathbf{H}_1 \cdot [\omega_2 \bar{\bar{\mu}}_2 - \omega_1 \bar{\bar{\mu}}_1^T] \cdot \mathbf{H}_2 - \mathbf{E}_1 \cdot [\omega_2 \bar{\bar{\varepsilon}}_2 - \omega_1 \bar{\bar{\varepsilon}}_1^T] \cdot \mathbf{E}_2 + (\omega_1 \mathbf{p}_1 \cdot \mathbf{E}_2 - \omega_2 \mathbf{p}_2 \cdot \mathbf{E}_1)\, dV, \quad (2)$$

where $\Sigma$ is an arbitrarily closed surface defining a volume $\Omega$.

### 1.3.2 QNM orthogonality

Lorentz reciprocity theorem can also be used to show that the QNMs satisfy an unconjugated orthogonality relation.

For that purpose, we set the second solution as the $n^{th}$ QNM of a system with a permittivity $\bar{\bar{\varepsilon}}$ and permeability $\bar{\bar{\mu}}$, that is $(\mathbf{E}_2, \mathbf{H}_2, \omega_2, \mathbf{p}_2, \bar{\bar{\varepsilon}}_2, \bar{\bar{\mu}}_2) = (\tilde{\mathbf{E}}_n^{(R)}, \tilde{\mathbf{H}}_n^{(R)}, \tilde{\omega}_n^{(R)}, \mathbf{0}, \bar{\bar{\varepsilon}}, \bar{\bar{\mu}})$, whereas the first solution corresponds to the $m^{th}$ QNM of a system with a permittivity $\bar{\bar{\varepsilon}}^T$ and permeability $\bar{\bar{\mu}}^T$, that is $(\mathbf{E}_1, \mathbf{H}_1, \omega_1, \mathbf{p}_1, \bar{\bar{\varepsilon}}_1, \bar{\bar{\mu}}_1) = (\tilde{\mathbf{E}}_m^{(L)}, \tilde{\mathbf{H}}_m^{(L)}, \tilde{\omega}_m^{(L)}, \mathbf{0}, \bar{\bar{\varepsilon}}^T, \bar{\bar{\mu}}^T)$. QNMs have divergent fields far away from the resonator and their treatment in a "generalized" Hilbert space requires regularization. We chose to regularize QNM with perfectly matched layers (PMLs) [2]. Thus, we choose the integral domain $\Omega$ to be the whole space including the PML regions so that Σ represents the outer surfaces of PMLs, which are made of perfect electric/magnetic conductors in practice. Perfect electric/magnetic conductors impose zero tangential components of electric/magnetic fields on Σ. As a result, the surface integral on the left-hand-side of Eq. (2) vanishes [3], and this leads to a simple volume integral relation

$$\iiint_\Omega \tilde{\mathbf{H}}_m^{(L)} \cdot [\tilde{\omega}_n \bar{\bar{\mu}}(\tilde{\omega}_n) - \tilde{\omega}_m \bar{\bar{\mu}}(\tilde{\omega}_m)] \cdot \tilde{\mathbf{H}}_n^{(R)} - \tilde{\mathbf{E}}_m^{(L)} \cdot [\tilde{\omega}_n \bar{\bar{\varepsilon}}(\tilde{\omega}_n) - \tilde{\omega}_m \bar{\bar{\varepsilon}}(\tilde{\omega}_m)] \cdot \tilde{\mathbf{E}}_n^{(R)} dV = 0. \quad (3)$$

Note that, for nonreciprocal systems, $\tilde{\mathbf{H}}_n^{(L)} \neq \tilde{\mathbf{H}}_n^{(R)}$ and $\tilde{\mathbf{E}}_n^{(L)} \neq \tilde{\mathbf{E}}_n^{(R)}$ , but they share the same eigenvalues, that is $\tilde{\omega}_n^{(R)} = \tilde{\omega}_n^{(L)} = \tilde{\omega}_n$ [5]. In the following part of the document, we refer to $(\tilde{\mathbf{H}}_m^{(L)}, \tilde{\mathbf{E}}_m^{(L)})$ as the left QNM and $(\tilde{\mathbf{H}}_n^{(R)}, \tilde{\mathbf{E}}_n^{(R)})$ as the right QNM.

Equation (3) is the orthogonality relation. Note that the permittivity and permeability are those of the actual (real) system without transposition. However, the orthogonality requires two fully independent computations, one for the real system QNMs and a second one for the transposed system. This is not the case for reciprocal materials.

### 1.3.3 The modal excitation coefficients and mode volumes

With the knowledge of the orthogonality of QNMs, we now discuss how to compute the QNM excitation coefficients. We consider a single electric dipole $\mathbf{p}$ located at $\mathbf{r}_0$ in the vicinity of a resonator. The dipole radiates a field $(\mathbf{E}, \mathbf{H})$ at the real frequency $\omega$. Owning to the completeness [3], we are able to expand the field everywhere in the PMLized space

$$\mathbf{E}(\mathbf{r}, \omega) = \sum_m \alpha_m(\omega) \tilde{\mathbf{E}}_m^{(R)}(\mathbf{r}), \quad (4)$$

where $\alpha_m$ are complex coefficients to be determined.

By applying Eq. (2) to the total field $(\mathbf{E}, \mathbf{H})$ created by the dipole at the frequency $\omega$ and to the $n^{th}$ left QNM $(\tilde{\mathbf{E}}_n^{(L)}, \tilde{\mathbf{H}}_n^{(L)})$, and using the modal expansion of Eq. (4), we obtain a linear system of equations $\sum_m B_{nm}(\omega)\alpha_m(\omega) = -\omega \mathbf{p} \cdot \tilde{\mathbf{E}}_n^{(L)}(\mathbf{r}_0)$, where the frequency-dependent coefficient $B_{nm}$ is given by

$$B_{nm} = \iiint_\Omega \tilde{\mathbf{H}}_n^{(L)} \cdot [\tilde{\omega}_n \bar{\bar{\mu}}(\tilde{\omega}_n) - \omega \bar{\bar{\mu}}(\omega)] \cdot \tilde{\mathbf{H}}_m^{(R)} - \tilde{\mathbf{E}}_n^{(L)} \cdot [\tilde{\omega}_n \bar{\bar{\varepsilon}}(\tilde{\omega}_n) - \omega \bar{\bar{\varepsilon}}(\omega)] \cdot \tilde{\mathbf{E}}_m^{(R)} dV. \quad (5)$$

In the absence of dispersion, all the non-diagonal terms are vanished according to Eq. (3) and the expansion coefficient is analytically given by

$$\alpha_n(\omega) = -\frac{\omega \mathbf{p} \cdot \tilde{\mathbf{E}}_n^{(L)}(\mathbf{r}_0)}{(\omega - \tilde{\omega}_n) \iiint_\Omega \tilde{\mathbf{E}}_n^{(L)} \cdot \bar{\bar{\varepsilon}} \cdot \tilde{\mathbf{E}}_n^{(R)} - \tilde{\mathbf{H}}_n^{(L)} \cdot \bar{\bar{\mu}} \cdot \tilde{\mathbf{H}}_n^{(R)} dV}. \quad (6)$$

In the general case of dispersive media, the off-diagonal coefficients $B_{nm}$ are not equal to zero. One thus has to solve a linear system of equations to obtain $\alpha_m$. To guarantee a safe and accurate numerical implementation, we note that $B_{nm}$ is null for $\omega = \tilde{\omega}_m$, and we introduce $A_{nm} = (\omega - \tilde{\omega}_m)^{-1} B_{nm}$, so that the linear system of equations can be rewritten as

$$\sum_m A_{nm}(\omega) x_m(\omega) = -\omega \mathbf{p} \cdot \tilde{\mathbf{E}}_n^{(L)}, \quad (7)$$

where the unknowns are now $x_m(\omega) = (\omega - \tilde{\omega}_m)\alpha_m(\omega)$.

The linear system Eq. (7) is not diagonal, and it is not possible in general to derive a closed-form expression. However, we can still show that $\alpha_m$ has a pole and use this property to derive an approximate analytical expression. For $\omega \approx \tilde{\omega}_n$, since $A_{nm}(\tilde{\omega}_n) = 0$ for $n \neq m$ according to Eq. (3), we therefore find

$$\alpha_n(\omega) \approx -\frac{\omega \mathbf{p} \cdot \tilde{\mathbf{E}}_n^{(L)}(\mathbf{r}_0)}{(\omega - \tilde{\omega}_n) \iiint_\Omega \tilde{\mathbf{E}}_n^{(L)} \cdot \frac{\partial(\omega \bar{\bar{\varepsilon}})}{\partial \omega} \cdot \tilde{\mathbf{E}}_n^{(R)} - \tilde{\mathbf{H}}_n^{(L)} \cdot \frac{\partial(\omega \bar{\bar{\mu}})}{\partial \omega} \cdot \tilde{\mathbf{H}}_n^{(R)} dV}. \quad (8)$$

We thus obtain an approximate closed-form expression for $\alpha_n$ valid in the vicinity of $\tilde{\omega}_n$. To derive Eqs. (6-8), we have strictly followed the approach in [2]. The approximate Eq. (8) has been successfully used in many earlier works starting with [2]. It is convenient, but it could also be set into a rigorous format. Using auxiliary fields to linearize the nonlinear eigenvalue problem of Eq. (1) (which is nonlinear in $\omega$ for dispersive material), we may derive an exact analytical expression for the excitation coefficients. This machinery that is described in [3] is not provided here for the sake of simplicity.

### 1.3.4 Mode volume

The Purcell factor is the modification of the spontaneous decay rate due to the resonance. Following [2], it is easily shown that the non-Hermitian Purcell factor $F_n$ of the resonance $\tilde{\mathbf{E}}_n^{(R)}$ (i.e. the) is

$$F_n = \frac{3}{4\pi^2}\left(\frac{\lambda_n}{n_b}\right)^3 \mathrm{Re}\left(\frac{Q_n}{\tilde{V}_n}\right) \frac{\Omega_n^2}{\omega^2} \frac{\Omega_n^2}{4Q_m^2(\omega - \Omega_n)^2 + \Omega_n^2}\left[1 + 2Q_n \frac{\omega - \Omega_n}{\Omega_n} \frac{\mathrm{Im}(\tilde{V}_n)}{\mathrm{Re}(\tilde{V}_n)}\right], \quad (9)$$

and that the mode volume is given

$$\tilde{V}_n = \frac{\iiint_\Omega \tilde{\mathbf{E}}_n^{(L)} \cdot \frac{\partial(\omega\bar{\bar{\varepsilon}})}{\partial\omega} \cdot \tilde{\mathbf{E}}_n^{(R)} - \tilde{\mathbf{H}}_n^{(L)} \cdot \frac{\partial(\omega\bar{\bar{\mu}})}{\partial\omega} \cdot \tilde{\mathbf{H}}_n^{(R)} dV}{2\varepsilon(\mathbf{r}_0)\left[\mathbf{p}\cdot\tilde{\mathbf{E}}_n^{(L)}\right]\left[\mathbf{p}\cdot\tilde{\mathbf{E}}_n^{(R)}\right]}. \quad (10)$$

The Purcell factor takes a form similar to that of the classical expression derived in a Hermitian framework, except for the last bracketed term. $Q_n = \Omega_n/\Gamma_n$ is the quality factor of the mode with complex eigenfrequency $\tilde{\omega}_n = \Omega_n - i\Gamma_n/2$. $n_b$ is the background refractive index at the dipole position. $\lambda_n = 2\pi c/\Omega_n$ is the resonance wavelength.

We, therefore, can define the normalization factor as

$$QN = \iiint_\Omega \tilde{\mathbf{E}}_n^{(L)} \cdot \frac{\partial(\omega\bar{\bar{\varepsilon}})}{\partial\omega} \cdot \tilde{\mathbf{E}}_n^{(R)} - \tilde{\mathbf{H}}_n^{(L)} \cdot \frac{\partial(\omega\bar{\bar{\mu}})}{\partial\omega} \cdot \tilde{\mathbf{H}}_n^{(R)} dV. \quad (11)$$

Equation (10) is very similar to that derived for reciprocal resonators in [2]. Note that the expression is unchanged by scaling either $[\tilde{\mathbf{E}}_n^{(L)}, \tilde{\mathbf{H}}_n^{(L)}]$ or $[\tilde{\mathbf{E}}_n^{(R)}, \tilde{\mathbf{H}}_n^{(R)}]$ by any multiplicative factor, as expected.

Note that Eq. (10) is identical to the expression given at the end of the 'Summary and Perspectives' Section in [6]; the present document thus provides a demonstration of the expression that was not demonstrated in [6] for compactness.

# 2. THE PERMITTIVITY AND PERMEABILITY OF YTTRIUM IRON GARNET

In this toolbox, we consider a simple case of a 2D Yttrium iron garnet (YIG) wire in a homogenous background. The users may easily extend it for studying resonators with other materials, dimensions, and geometries following the steps given in the next Tutorial section.

Under an external dc magnetic field along the $z$ direction, the YIG exhibits a strong gyromagnetic anisotropy, with the relative permeability tensor taking the form

$$\bar{\bar{\mu}} = \begin{bmatrix} \mu_r & -i\kappa_r & 0 \\ i\kappa_r & \mu_r & 0 \\ 0 & 0 & \mu_\infty \end{bmatrix} \mu_0, \quad (12)$$

where the matrix elements $\mu_r$ and $\kappa_r$ are determined by the Landau-Lifshitz-Gilbert equation and can be expressed as [7]

$$\mu_r = \mu_\infty \left[1 + \frac{(\omega_H - i\alpha\omega)\omega_M}{(\omega_H - i\alpha\omega)^2 - \omega^2}\right], \quad (13)$$

and

$$\kappa_r = \frac{\mu_\infty \omega\omega_M}{(\omega_H - i\alpha\omega)^2 - \omega^2}, \quad (14)$$

where $\mu_\infty = 1$, $\alpha$ is the damping parameter, $\omega_H = \gamma H_0$, $\omega_M = \gamma M_s$ with $\gamma$ and $M_s$ being the gyromagnetic ratio and saturated magnetization, respectively, $H_0$ being the external static magnetic field which leads to broken time-reversal symmetry. On the other hand, the YIG permittivity is taken as a constant scalar $\bar{\bar{\varepsilon}} = \varepsilon_r\varepsilon_0 \,\mathrm{diag}(1,1,1)$.

At last, note the simple relation [8]

$$\bar{\bar{\mu}}^{-1} = \begin{bmatrix} d & -ib & 0 \\ ib & d & 0 \\ 0 & 0 & \mu_\infty^{-1} \end{bmatrix} \mu_0^{-1}, \quad (15)$$

where $d = \mu_r/(\mu_r^2 - \kappa_r^2)$ and $b = -\kappa_r/(\mu_r^2 - \kappa_r^2)$, which may help to derive the weak form expression later.

# 3. A TUTORIAL: COMPUTING QNMS FOR A YIG WIRE IN AIR

In this part, we provide details on how to compute the QNMs of a classical nonreciprocal resonator, a YIG wire in a homogenous background. Note that in QNMEig, there are already several well-documented examples for computing modes for dispersive resonators. However, they are all limited to systems with reciprocal materials, and **QNMnonreciprocal_resonators** can be considered as an extension of the **QNMEig** for non-reciprocal systems.

## 2.1 Modeling Instructions

Open COMSOL Multiphysics. From its **File** menu, choose **New**.

NEW

In the **New** window, click **Model Wizard**.

MODEL WIZARD

Since we need to compute both the left and right QNMs, unlike the other models in **QNMEig**, here we need two emws and two weak form PDEs.

1. In the **Model Wizard** window, click **2D**.
2. In the **Select physics** tree, select Radio **Frequency->Electromagnetic Waves, Frequency Domain (emw)**.
3. Click **Add**.
4. In the **Select physics** tree, select **Mathematics->PDE Interfaces, Weak Form PDE (w)**.
5. Click **Add**.
6. In the **Select physics** tree, select **Radio Frequency->Electromagnetic Waves, Frequency Domain (emw)**.
7. Click **Add**.
8. In the **Select physics** tree, select **Mathematics->PDE Interfaces, Weak Form PDE (w)**.
9. Click **Add**.
10. Click **Study**
11. In the **Select Study** tree, select Preset Studies for **Selected Physics Interfaces>Eigenfrequency**
12. Click **Done**

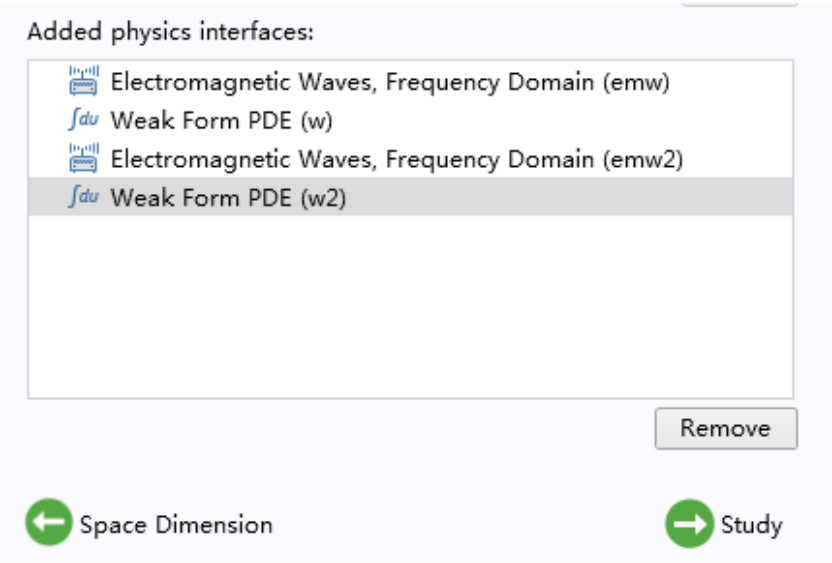

Here the ‘emw2’ and ‘w2’ are built for computing the left QNM.

GLOBAL DEFINITIONS

Define the model geometric and material parameters.

Parameters 1

1. In the **Model Builder** window, under **Global Definitions** click **Parameters 1**.
2. In the **Settings** window for the **Parameters**, locate the **Parameters** section.
3. In the table, enter the following settings:

| Name | Expression | Description |
|---|---|---|
| Lair | a*1.5 | Geom: air background width |
| r | 0.35*a | Geom: wire radius |
| Lpml | a/4 | Geom: PML thickness |
| a | 26[mm] | Geom: length of the air domain |
| epsrinf | 15 | Material: permittivity of YIG |
| murinf | 1 | Material: mu_inf given in section 2 |
| epsilonb | 1 | Material: permittivity of the background medium |
| lambda_pml | c_const/freqg | Material: typical absorbing wavelength of PMLs |
| freqg | 8.8466 [GHz] | Freq: frequency to search for QNMs |
| Nomega | (freqg*2*pi) | Freq: normalization factor for the weak form |

The Nomega can be set as an arbitrary value but should be of the same order of freqg*2*pi. It is used in the auxiliary-field equation of the Weak Form PDE module (see later) to make the two equations of the quadratic polynomial eigenproblem [3] have similar magnitudes, thereby increasing the numerical stabilities.

Parameters 2

4. In the **Model Builder** window, right-click **Global Definitions** and select **Parameters**.
5. Type YIG materials in the **Label** text field.
6. In the **Settings** window for **Parameters 2**, locate the **Parameters** section.
7. In the table, enter the following settings:

| Name | Expression | Description |
|---|---|---|
| omegam1 | 175[mT]*gamma1 | Material: omega_m given in section 2 |
| omega01 | gamma1*Hs1*mu0_const | Material: omega_0 given in section 2 |
| Hs1 | 900[Oe] | Material: H0 given in section 2 |
| gamma1 | 2*pi*28[GHz/T] | Material: gamma given in section 2 |
| alpha1 | 3e-4 | Material: alpha given in section 2 |

This table gives the parameters which are used to compute the $\bar{\bar{\mu}}$ of YIG.

GEOMETRY

The geometry consists of a YIG wire in air background, surrounded by a PML.

Circle 1

1. In the **Geometry** toolbar, click **Circle**.
2. In the **Settings** window for **Circle**, locate the **Size and Shape** section.
3. In the **Radius** text field, type r.

Rectangle 1

4. In the **Geometry** toolbar, click **Rectangle**.
5. In the **Settings** window for **Rectangle**, locate the **Size and Shape** section.
6. In the **Width** text field, type Lair+Lpml*2.
7. In the **Height** text field, type Lair+Lpml*2.
8. Locate the **Position** section. From the **Base** list, select **Center**.
9. Locate the **Layers** section. In the **Layer 1** text field, type Lpml.
10. Choose **Layers to the left**, **Layers to the right**, **Layers on bottom**, and **Layers on top** checkboxes.
11. Click the **Build All Objects**.

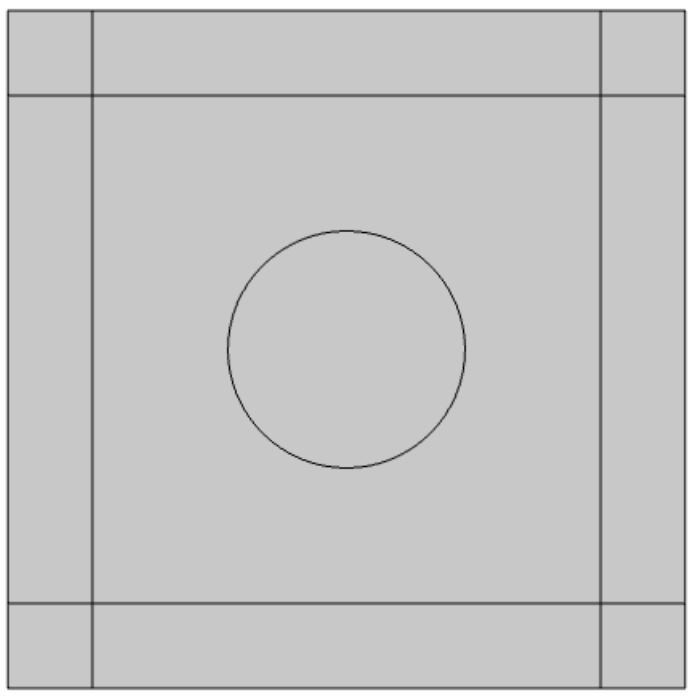

The geometry of the system: a YIG wire in air background, surrounded by a PML.

DEFINITIONS

Define PML domains and PML types. Define the resonator and background. Define different variables used for normalizing the QNMs.

Explicit 1

1. In the **Definitions** toolbar, click **Explicit**.
2. In the **Settings** window for **Explicit**, type PML in the **Label** text field.
3. Set Domains 1, 2, 3, 4, 6, 7, 8, 9.

Explicit 2

4. In the **Definitions** toolbar, click **Explicit**.
5. In the **Settings** window for Explicit, type sca in the **Label** text field.
6. Set Domain 10 only.

Explicit 3

7. In the **Definitions** toolbar, click **Explicit**.
8. In the **Settings** window for **Explicit**, type Air background and its attached PML in the **Label** text field.
9. Set Domains 1, 2, 3, 4, 5, 6, 7, 8, 9.

Perfectly matched layers

We model the PML as a non-dispersive material.

10. In the **Definitions** toolbar, click **Perfectly Matched Layer**.
11. In the **Settings** window for **Perfectly Matched Layer**, locate the **Domain Selection** section.
12. From the **Selection** list, choose **PML**.
13. Locate the **Geometry** section. From the **Type** list, select **Cartesian**.
14. Locate the **Scaling** section. From the **Typical wavelength from** list, choose **User defined**.
15. In **Typical wavelength** text field, type lambda_pml.

Variables 1

Here we define the complex eigenfrequencies that will be used in the weak form later.

16. On the **Definitions** toolbar, click **Variables**.
17. In the **Settings** window for Variables, locate the **Geometric Entity Selection** section.
18. Locate the **Variables** section. In the table, enter the following settings:

| Name | Expression | Description |
|---|---|---|
| QNM_omega | emw.iomega/i | Complex eigenfrequency of E^L |
| QNM_omega2 | emw2.iomega/i | Complex eigenfrequency of E^R |

Variables 2

Just like in the post-processing step, we are going to use values of the auxiliary fields everywhere, thereby defining them to be 0 outside the resonator.

19. In the **Definitions** toolbar, click **Variables**.
20. In the **Settings** window for **Variables**, locate the **Geometric Entity Selection** section.
21. From the **Geometric entity level** list, choose **Domain**.
22. From the **Selection** list, choose **Air background and its attached PML**.
23. Locate the **Variables** section. In the table, enter the following settings:

| Name | Expression | Description |
| --- | --- | --- |
| M1x | 0[V/m^2] | Auxiliary field M1x in Air domain |
| M1y | 0[V/m^2] | Auxiliary field M1y in Air domain |
| N1x | 0[V/m^2] | Auxiliary field N1x in Air domain |
| N1y | 0[V/m^2] | Auxiliary field N1y in Air domain |

Variables 3

Define the analytical expression of $\partial(\omega\bar{\bar{\mu}})/\partial\omega\widetilde{\mathbf{H}}_n^{(R)}$. Its three components are given by dwudwH_x, dwudwH_y, and dwudwH_z. These values will be used later to compute $QN$.

24. In the **Definitions** toolbar, click **Variables**.
25. In the **Settings** window for Variables, locate the **Geometric Entity Selection** section.
26. From the **Geometric entity level** list, choose **Domain**.
27. From the **Selection** list, choose **sca**.
28. Locate the **Variables** section. In the table, enter the following settings:

| Name | Expression | Description |
| --- | --- | --- |
| fac1 | i*alpha1*w2+omega01+omegam1 | |
| fac3 | i*w2*omegam1 | |
| fac2 | i*alpha1*w2+omega01 | |
| lowduinvu | (w2^2-fac1^2)*(w2^2-fac2^2) | |
| fac4 | (2*omega01+omegam1)*w2*(1+alpha1^2) | |
| fac5 | i*alpha1*((1+alpha1^2)*w2^2-omega01*(omega01+omegam1)) | |
| upduinvu11 | omegam1*(fac4+fac5) | |
| upduinvu12 | i*omegam1*((1+alpha1^2)*w2^2+omega01*(omega01+omegam1)) | |
| dwudw11 | 1/w2+upduinvu11/lowduinvu | |
| dwudw12 | upduinvu12/lowduinvu | |
| dwudw21 | -upduinvu12/lowduinvu | |
| dwudw22 | 1/w2+upduinvu11/lowduinvu | |
| dwudwH_x | (dwudw11*cERx+dwudw12*cERy)/(-i)/mu0_const | |
| dwudwH_y | (dwudw21*cERx+dwudw22*cERy)/(-i)/mu0_const | |
| dwudwH_z | cERz/w2/(-i)/mu0_const | |

Variables 4

Define the analytical expression of $\partial(\omega\bar{\bar{\mu}})/\partial\omega\widetilde{\mathbf{H}}_n^{(R)}$ outside the resonator.

29. In the **Definitions** toolbar, click **Variables**.
30. In the **Settings** window for Variables, locate the **Geometric Entity Selection** section.
31. From the **Geometric entity level** list, choose **Domain**.
32. From the **Selection** list, choose **Air background and its attached PML**.
33. Locate the **Variables** section. In the table, enter the following settings:

| Name | Expression | Description |
| --- | --- | --- |
| dwudwH_x | HRx | |
| dwudwH_y | HRy | |
| dwudwH_z | HRz | |

Variables 5

Since we have changed the weak form, emw(2).Hx, emw(2).Hy, and emw(2).Hz no longer give the correct magnetic fields. HLx, HLy, and HLz are the correct magnetic field for the left QNMs. HRx, HRy, and HRz are the correct magnetic field for the right QNMs. We also give the curl of $\tilde{\mathbf{E}}_n^{(R)}$, cERx, cERy, and cERz.

34. On the **Definitions** toolbar, click **Variables**.
35. In the **Settings** window for Variables, locate the **Geometric Entity Selection** section.
36. Locate the **Variables** section. In the table, enter the following settings:

| Name | Expression | Description |
|---|---|---|
| w | emw.iomega/i | |
| w2 | emw2.iomega/i | |
| HLx | 1/(-i*w*mu0_const)*(M1x+emw.curlEx*invmuinf) | |
| HLy | 1/(-i*w*mu0_const)*(M1y+emw.curlEy*invmuinf) | |
| HLz | 1/(-i*w*mu0_const)*(emw.curlEz*invmuinf) | |
| invmuinf | (murinf)^(-1) | |
| cERx | emw2.curlEx | |
| cERy | emw2.curlEy | |
| cERz | emw2.curlEz | |
| HRx | 1/(-i*w2*mu0_const)*(N1x+emw2.curlEx*invmuinf) | |
| HRy | 1/(-i*w2*mu0_const)*(N1y+emw2.curlEy*invmuinf) | |
| HRz | 1/(-i*w2*mu0_const)*(emw2.curlEz*invmuinf) | |

Variables 6

Define the numerical expression of $\partial(\omega\bar{\bar{\mu}})/\partial\omega\tilde{\mathbf{H}}_n^{(R)}$. In contrast to **Variables 3**, here these values are computed numerically.

37. In the **Definitions** toolbar, click **Variables**.
38. In the **Settings** window for Variables, locate the **Geometric Entity Selection** section.
39. From the **Geometric entity level** list, choose **Domain**.
40. From the **Selection** list, choose **sca**.
41. Locate the **Variables** section. In the table, enter the following settings:

| Name | Expression | Description |
|---|---|---|
| mur_YIG | murinf+murinf*omegam1*(omega01+i*alpha1*w2)/((omega01+i*alpha1*w2)^2-w2^2) | |
| kappa_YIG | omegam1*w2/((omega01+i*alpha1*w2)^2-w2^2)*murinf | |
| dwkappa_YIG | d(kappa_YIG*w2,w2) | |
| dmur_YIG | d(mur_YIG*w2,w2) | |
| dmu11 | dmur_YIG | |
| dmu12 | dwkappa_YIG*i | |
| dmu21 | -dwkappa_YIG*i | |
| dmu22 | dmur_YIG | |
| duwHx | dmu11*HRx+dmu12*HRy | |
| duwHy | dmu21*HRx+dmu22*HRy | |
| duwHz | HRz | |

Variables 7

Define the numerical expression of $\partial(\omega\bar{\bar{\mu}})/\partial\omega\tilde{\mathbf{H}}_n^{(R)}$ outside the resonator.

42. In the **Definitions** toolbar, click **Variables**.
43. In the **Settings** window for Variables, locate the **Geometric Entity Selection** section.
44. From the **Geometric entity level** list, choose **Domain**.
45. From the **Selection** list, choose **Air background and its attached PML**.
46. Locate the **Variables** section. In the table, enter the following settings:

| Name | Expression | Description |
| --- | --- | --- |
| duwHx | HRx | |
| duwHy | HRy | |
| duwHz | HRz | |

ELECTROMAGNETIC WAVES, FREQUENCY DOMAIN (emw)

emw and w are used to solve for the left QNMs. Here we only consider the case where the electric field component of the QNM is polarized along $z$ axis.

1. In the **Settings** window for **Electromagnetic Waves, Frequency Domain**, locate the **Components** section.
2. From the **Electric field components solved for list**, choose **Out-of-plane vector**.
3. On the top of the **Modal Builder** window, click **Show** option (the one with an eye icon), select at least **Advanced Physics and Discretization options**.
4. In the **Physics** toolbar, select **Electromagnetic Waves, Frequency Domain (emw)**.
5. In the **Settings** window for **Electromagnetic Waves, Frequency Domain (emw)**, locate the Domain Selection section.
6. From the **Selection** list, choose **All Domains**.

Weak contribution 1

Here we define the weak contribution. Inside the resonator COMSOL solve the integration equation: $\int_V \nabla\times\mathbf{F}(\mathbf{r})\cdot\mu_\infty^{-1}\nabla\times\tilde{\mathbf{E}}_n^{(L)}(\mathbf{r})-\tilde{\omega}_n^2\mathbf{F}\cdot\varepsilon\tilde{\mathbf{E}}_n^{(L)}(\mathbf{r})+\nabla\times\mathbf{F}(\mathbf{r})\cdot\mathbf{M}\,(\mathbf{r})dV=0$ with $\mathbf{M}(\mathbf{r})=((\bar{\bar{\mu}}^T)^{-1}-\mu_\infty^{-1})\nabla\times\tilde{\mathbf{E}}_n^{(L)}(\mathbf{r})$. The first and second terms are defined in **Wave Equation, Electric**, whereas the third term is difined in **Weak contribution 1**.

7. In the **Physics** toolbar, from the **Domains** section list, choose **Weak Contribution**.
8. From the **Selection** list, choose **sca**.
9. In **Weak expression** text field, type -M1x*test(emw.curlEx)-M1y*test(emw.curlEy).

Wave Equation, Electric 1

The weak expression for the resonator domain.

10. In the **Settings** window for **Wave Equation, Electric 1**, locate the **Electric Displacement Field** section.
11. From the **Electric displacement field model** list, choose **Relative permittivity**.
12. From **Relative permittivity** list, choose **User defined**. In the text field, enter epsrinf.
13. In the **Settings** window for **Variables**, locate the **Magnetic Field** section.
14. From the **Constitutive relation**, choose **Relative permeability**.
15. From **Relative permeability** list, choose **User defined**. In the text field, enter murinf.
16. In the **Settings** window for **Variables**, locate the **Conduction Current** section.
17. From the **Electric conductivity**, choose **User defined**. In the text field, enter 0.

Wave Equation, Electric 2

The weak expression for the background domain.

18. In the **Physics** toolbar, from the **Domains** section list, choose **Wave Equation, Electric**.
19. In the **Settings** window for **Wave Equation, Electric 2**, locate **Domain Selection** section.
20. From the **Selection** list, choose **Air background and its attached PML**.
21. From the **Electric displacement field model** list, choose **Relative permittivity**.
22. From **Relative permittivity** list, choose **User defined**. In the text field, enter epsilonb.
23. In the **Settings** window for **Variables**, locate the **Magnetic Field** section.
24. From the **Constitutive relation**, choose **Relative permeability**.
25. From **Relative permeability** list, choose **User defined**. In the text field, enter 1.
26. In the **Settings** window for **Variables**, locate the **Conduction Current** section.
27. From the **Electric conductivity**, choose **User defined**. In the text field, enter 0.

Weak Form PDE (w)

The weak expression for the auxiliary fields M1x and M1y. $\mathbf{M}(\mathbf{r}) \cdot \mathbf{F} = ((\bar{\bar{\mu}}^T)^{-1} - \mu_\infty^{-1})\nabla \times \tilde{\mathbf{E}}_n^{(L)}(\mathbf{r}) \cdot \mathbf{F}$.

1. In the **Physics** toolbar, select **Weak Form PDE**.
2. In the **Settings** window for **Weak Form PDE**, locate the **Domain Selection** section.
3. From the **Selection** list, choose: **sca**.
4. Locate the **Discretization** section. From the **Shape function type** list, choose **Lagrange**.
5. Locate the **Dependent Variables** section. In the **Field name** text field, enter M1. In the **Number of dependent** variables text field, enter 2. In the **Dependent variables** text field, enter M1x, M1y.

Weak Form PDE 1

6. In the **Modal Builder** window, under the **Weak Form PDE (w)** module, click **Weak Form PDE 1**.
7. In the **Settings** window for **Weak Form PDE 1**. Locate the **Weak Expressions** section, enter the following expression:

(test(M1x)*emw.curlEx+test(M1y)*emw.curlEy)*(omegam1*(i*alpha1*QNM_omega+omega01+omegam1))/nomega^2+(test(M1x)*emw.curlEy-test(M1y)*emw.curlEx)*(-i)*(QNM_omega*omegam1)/nomega^2-(test(M1x)*M1x+test(M1y)*M1y)*((1+alpha1^2)*QNM_omega^2-2*i*alpha1*QNM_omega*(omega01+omegam1)-(omega01+omegam1)^2)*murinf/nomega^2

ELECTROMAGNETIC WAVES, FREQUENCY DOMAIN (emw2)

emw and w are used to solve for the right QNMs.

Repeat what has been done in **ELECTROMAGNETIC WAVES, FREQUENCY DOMAIN (emw1)** but replace all the M1x and M1y with N1x and N1y. Note that $\mathbf{N}(\mathbf{r}) = (\bar{\bar{\mu}}^{-1} - \mu_\infty^{-1})\nabla \times \tilde{\mathbf{E}}_n^{(R)}(\mathbf{r})$, which is different from $\mathbf{M}(\mathbf{r})$.

Weak Form PDE (w2)

Repeat what has been done in **Weak Form PDE (w)** but replace all the M1x and M1y with N1x and N1y and in the **Settings** window for **Weak Form PDE 1**. Locate the **Weak Expressions** section, enter the following expression:

(test(N1x)*emw2.curlEx+test(N1y)*emw2.curlEy)*(omegam1*(i*alpha1*QNM_omega2+omega01+omegam1))/nomega^2+(test(N1x)*emw2.curlEy-test(N1y)*emw2.curlEx)*(i)*(QNM_omega2*omegam1)/nomega^2-(test(N1x)*N1x+test(N1y)*N1y)*((1+alpha1^2)*QNM_omega2^2-2*i*alpha1*QNM_omega2*(omega01+omegam1)-(omega01+omegam1)^2)*murinf/nomega^2

Mesh 1

Free Triangular 1

1. In the **Model Builder** window, right-click **Mesh 1** and choose **Free Triangular**.
2. In the **Settings** window for **Free Triangular** locate the **Domain Selection** section.
3. From the **Geometric** entity level list, choose **Domain**.
4. From the **Selection** list, choose **sca**.
5. Right-click **Free Triangular 1** and choose **Size**.
6. In the **Settings** window for **Size 1**, Locate the **Element Size** section and choose **Custom**.
7. From the **Geometric entity level** list, choose **Domain**.
8. From the **Selection** list, choose **sca**.
9. Locate the **Element Size** section, In the **Maximum element size** text field, type r/10.

Free Triangular 2

10. In the **Model Builder** window, right-click **Mesh 1** and choose **Free Triangular**.
11. In the **Settings** window for **Free Triangular** locate the **Domain Selection** section.
12. From the **Geometric** entity level list, choose **Domain**.
13. Select Domain 5 only.
14. Right-click **Free Triangular 1** and choose **Size**.
15. In the **Settings** window for **Size 1**, locate the **Element Size** section and choose **Custom**.
16. From the **Geometric entity level** list, choose **Domain**.
17. Select Domain 5 only.
18. Locate the **Element Size** section, In the **Maximum element size** text field, type r/2.

Mapped 1

19. In the **Model Builder** window, right-click **Mesh 1** and choose **Mapped**.
20. In the **Settings** window for **Mapped** locate the **Domain Selection** section.
21. From the **Geometric** entity level list, choose **Remaining**.
22. Right-click **Mapped 1** and choose **Distribution**.
23. In the **Settings** window for **Distribution 1**, locate the **Distribution** section.
24. From the **Distribution** type list, choose **Fixed number of elements**.
25. In the **Number of elements** field, type 5.
26. Click **Build All**.

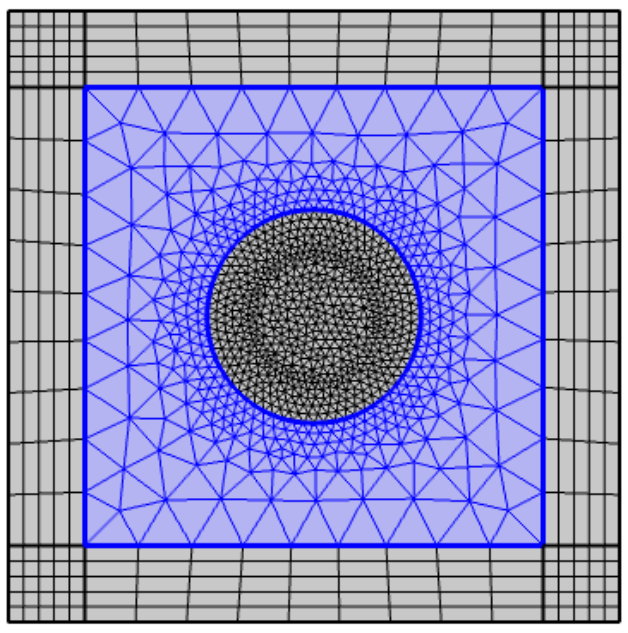

The mesh of the system.

Study 1

Search for the left QNMs.

1. In the Model Builder window, click **Study 1>>Step 1: Eigenfrequency**.
2. Activate the **Desired number of eigenfrequencies** check box. Type the number of modes.
3. From the **Unit** list, choose **GHz**.
4. In the **Search for eigenfrequencies around**, type freqg.
5. Locate the **Physics and Variables Selection** section, clear the **ELECTROMAGNETIC**

**WAVES, FREQUENCY DOMAIN 2 (emw2)** and **Weak Form PDE 2 (w2)** checkboxes.

6. Click **Compute**.

Study 2

Search for the right QNMs.

7. In the **Study** toolbar, select **Add study**.
8. In the **Add study** window, click **General Studies>> Eigenfrequency**.
9. In the **Model Builder** window, click **Study 2>>Step 1: Eigenfrequency**.
10. Activate the **Desired number of eigenfrequencies** check box. Type the number of modes and make sure it is equal to that in **Study 1**.
11. From the **Unit** list, choose **GHz**.
12. In the **Search for eigenfrequencies around**, type freqg.
13. Locate the **Physics and Variables Selection** section, clear the **ELECTROMAGNETIC WAVES, FREQUENCY DOMAIN (emw)** and **Weak Form PDE (w)** checkboxes.
14. Click **Compute**.

Data Sets

Join 1

Because we need both the left QNMs computed with Study 1 and the right QNM computed with Study 2, we need to join two data sets.

1. In the **Study** toolbar, select **More Data Sets>>Base Data Det>>Join**.
2. In the **Model Builder** window, click **Join 1.**
3. In the **Settings** window for **Join 1**, locate the **Data 1** section.
4. From the **Data** list, choose **Study 1/Solution 1 (sol1)**.
5. From the **Solutions** list, choose **One**.
6. From the **Eigenfrequency list**, choose a mode which we are concerned about.
7. In the **Settings** window for **Join 1**, locate the **Data 2** section.
8. From the **Data** list, choose **Study 2/Solution 2 (sol2)**.
9. From the **Solutions** list, choose **One**.
10. From the **Eigenfrequency list**, choose the mode which has the same frequency as that in **Data 1**.
11. In the **Settings** window for **Join 1**, locate the **Combination** section.
12. From the **Combination** list, choose **Explicit**.

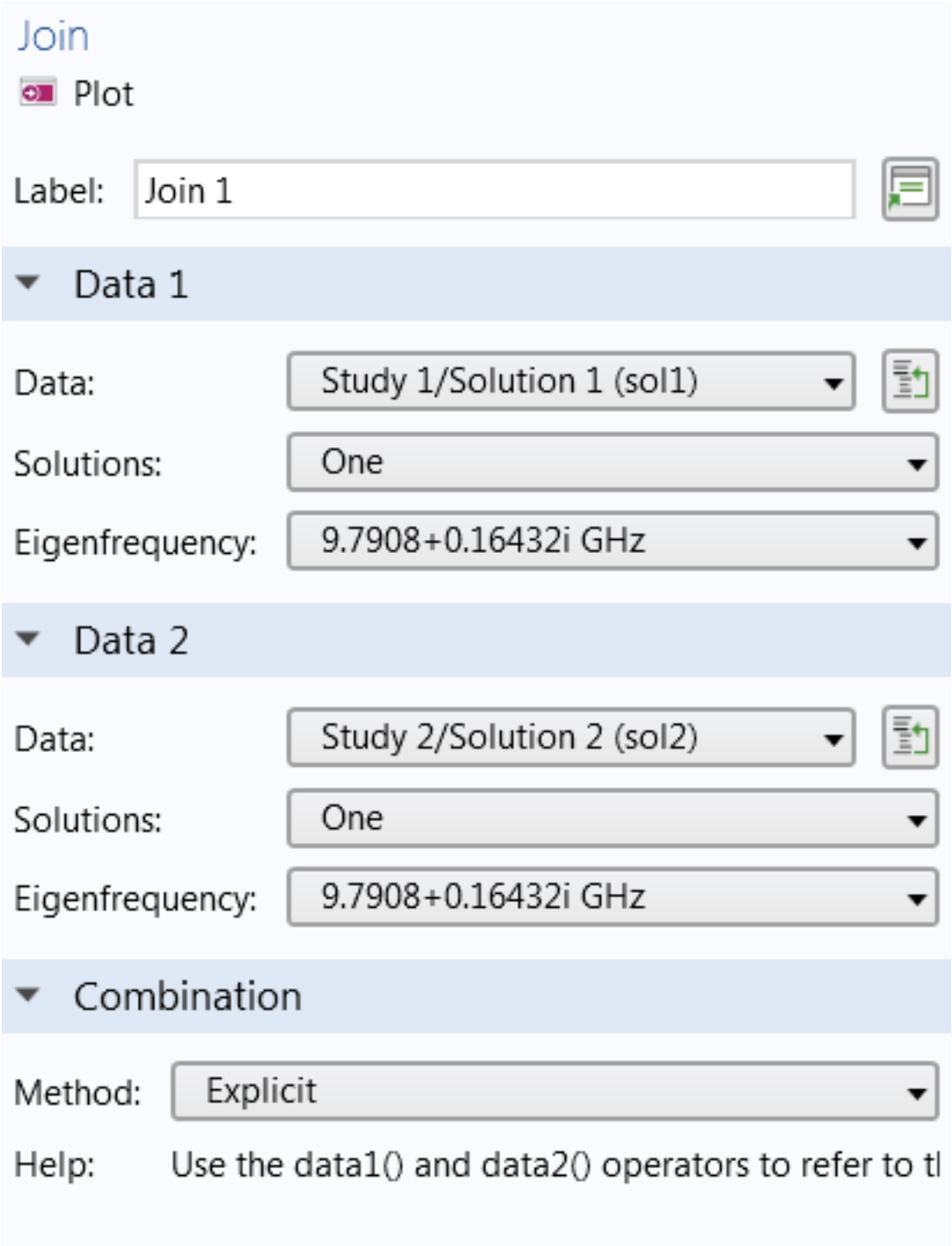

Derived Values

Surface integral
Evaluate the $QN$ defined in Eq. (11).

1. In the **Study toolbar**, select **More Derived Values>>Integration>>Surface integration**.
2. In the **Model Builder** window, click **Surface integration 1**.
3. From the **Data set** list, choose **Join 1**.
4. Locate the **Expressions** section. In the table, enter the following settings:

| Expression | Description |
|---|---|
| (data1(emw.Ex)*data2(emw2.Dx)+data1(emw.Ey)*data2(emw2.Dy) +data1(emw.Ez)*data2(emw2.Dz))*data1(pml1.detInvT) | QN_E |
| -(data1(HLx)*data2(dwudwH_x)+data1(HLy)*data2(dwudwH_y) +data1(HLz)*data2(dwudwH_z))*data1(pml1.detInvT)*mu0_const | QN_H |
| (data1(emw.Ex)*data2(emw2.Dx)+data1(emw.Ey)*data2(emw2.Dy) +data1(emw.Ez)*data2(emw2.Dz))*data1(pml1.detInvT)-(data1(HLx)*data2(dwudwH_x)+data1(HLy)*data2(dwudwH_y) +data1(HLz)*data2(dwudwH_z))*data1(pml1.detInvT)*mu0_const | QN=QN_E+QN_H |
| -(data1(HLx)*data2(duwHx)+data1(HLy)*data2(duwHy) +data1(HLz)*data2(duwHz))*mu0_const*data1(pml1.detInvT) | QN_H (Method 2) |

5. Click **Evaluate>>New table**.

2D Plot Group

Plot the map for $\mathrm{Re}(\tilde{\mathbf{E}}_m^{(R)} \cdot \tilde{\mathbf{E}}_m^{(L)}/QN)$.

1. In the **Study** toolbar, select **2D Plot Group.**
2. From the **Data set** list, choose **Join 1**.
3. In the **Model Builder** window, right-click **2D Plot Group 2**, select **Surface**.

Surface 1

4. In the **Model Builder** window, click **Surface.**
5. Locate the **Expressions** section. In the table, enter the following settings: real(data1(emw.Ez)*data2(emw2.Ez)/X) with X being the value of QN=QN_E+QN_H computed in the previous step.

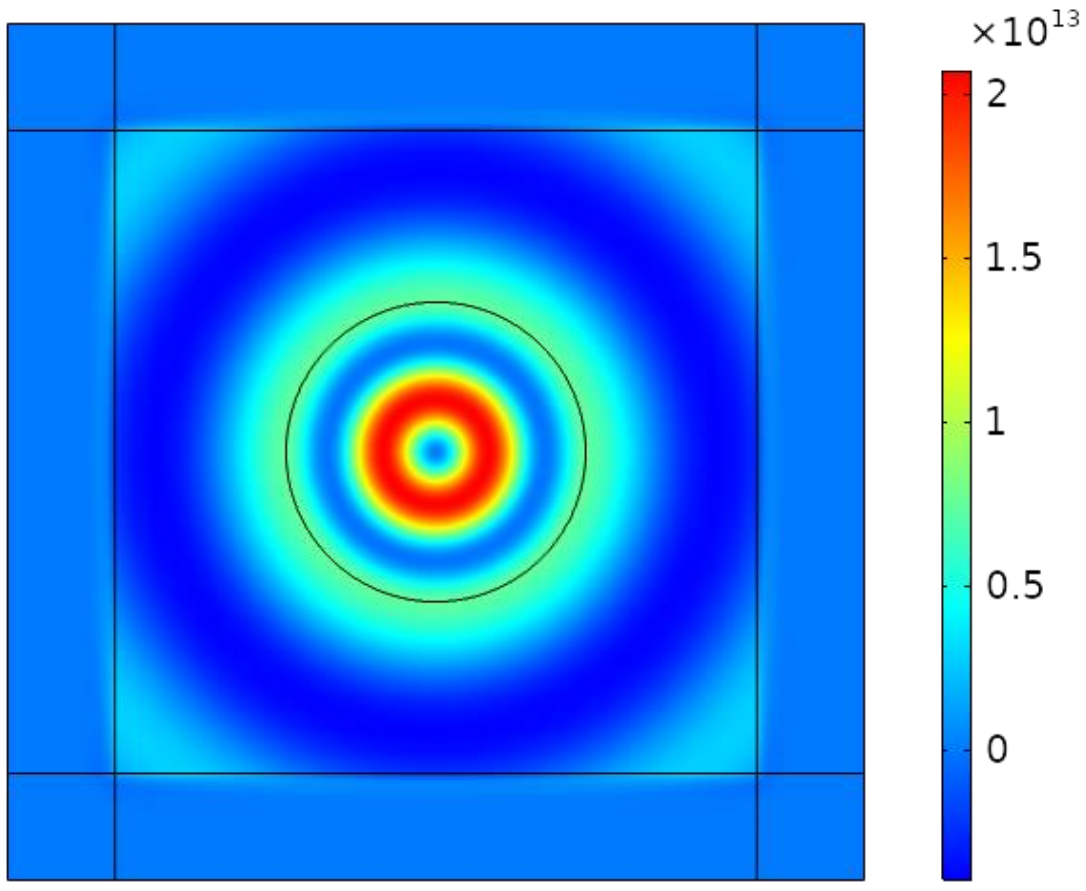

$\mathrm{Re}(\tilde{\mathbf{E}}_m^{(R)} \cdot \tilde{\mathbf{E}}_m^{(L)}/QN)$

## 4. REFERENCES

1. Bai, Q.; Perrin, M.; Sauvan, C.; Hugonin, J. P.; Lalanne, P. "Efficient and intuitive method for the analysis of light scattering by a resonant nanostructure", *Opt. Express* **2013**, 21, 27371.
2. Sauvan, C.; Hugonin, J. P.; Maksymov, I. S.; Lalanne, P. "Theory of the spontaneous optical emission of nanosize photonic and plasmon resonators", *Phys. Rev. Lett.* **2013**, 110, 237401.
3. Yan, W.; Faggiani, R.; Lalanne, P. "Rigorous modal analysis of plasmonic nanoresonators", *Phys. Rev. B* **2018**, 97, 205402.
4. Vial, B.; Nicolet, A.; Zolla, F.; Commandré, M. "Quasimodal expansion of electromagnetic fields in open two-dimensional structures", *Phys. Rev. A* **2014**, 89, 023829.
5. Pick, A.; Zhen, B.; Miller O.D.; Hsu C.W.; Hernandez F.; Rodriguez A.W.; Soljačić M.; Johnson S.G. "General Theory of Spontaneous Emission Near Exceptional Points", *Opt. Express* **2017**, 25, 12325.
6. Wu, T.; Gurioli, M.; Lalanne, P. "Nanoscale Light Confinement: the Q's and V's", *ACS Photonics* **2021**, (https://doi.org/10.1021/acsphotonics.1c00336).
7. Rameshti, B. Z.; Bauer, G. E. "Indirect coupling of magnons by cavity photons", *Phys. Rev. B* **2018**, 97, 014419.
8. Yang, B.; Wu, T; Zhang, X. "Topological properties of nearly flat bands in two-dimensional photonic crystals", *J. Opt. Soc. Am. B* **2017**, 34, 831.